\documentclass[aps,prb,twocolumn,superscriptaddress,amsmath]{revtex4-1}

\usepackage{graphicx}

\begin{document}


\title{Charge imbalance in superconductors in the low-temperature limit}


\author{F. H\"ubler}
\affiliation{Institut f\"ur Nanotechnologie, Karlsruher Institut f\"ur Technologie, Karlsruhe, Germany}
\affiliation{Center for Functional Nanostructures, Karlsruher Institut f\"ur Technologie, Karlsruhe, Germany}
\affiliation{Institut f\"ur Festk\"orperphysik, Karlsruher Institut f\"ur Technologie, Karlsruhe, Germany}
\author{J. Camirand Lemyre}
\affiliation{Institut f\"ur Nanotechnologie, Karlsruher Institut f\"ur Technologie, Karlsruhe, Germany}
\author{D. Beckmann}
\email[e-mail address: ]{detlef.beckmann@kit.edu}
\affiliation{Institut f\"ur Nanotechnologie, Karlsruher Institut f\"ur Technologie, Karlsruhe, Germany}
\affiliation{Center for Functional Nanostructures, Karlsruher Institut f\"ur Technologie, Karlsruhe, Germany}
\author{H. v. L\"ohneysen}
\affiliation{Center for Functional Nanostructures, Karlsruher Institut f\"ur Technologie, Karlsruhe, Germany}
\affiliation{Institut f\"ur Festk\"orperphysik, Karlsruher Institut f\"ur Technologie, Karlsruhe, Germany}
\affiliation{Physikalisches Institut, Karlsruher Institut f\"ur Technologie, Karlsruhe, Germany}

\date{\today}

\begin{abstract}
We explore charge imbalance in mesoscopic normal-metal/superconductor multiterminal structures at very low temperatures. The investigated samples, fabricated by e-beam lithography and shadow evaporation, consist of a superconducting aluminum bar with several copper wires forming tunnel contacts at different distances from each other. We have measured in detail the local and non-local conductance of these structures as a function of the applied bias voltage $V$, the applied magnetic field $B$, the temperature $T$ and the contact distance $d$. From these data the charge-imbalance relaxation length $\lambda_{Q^*}$ is derived. The bias-resolved measurements show a transition from dominant elastic scattering close to the energy gap to an inelastic two-stage relaxation at higher bias. We observe a strong suppression of charge imbalance with magnetic field, which can be directly linked to the pair-breaking parameter. In contrast, practically no temperature dependence of the charge-imbalance signal was observed below 0.5~K. These results are relevant for the investigation of other non-local effects such as crossed Andreev reflexion and spin diffusion.

\end{abstract}

\pacs{74.40.Gh, 74.25.F-}

\maketitle


\section{Introduction}
Non-equilibrium phenomena in superconductors have been investigated intensively since the 1970s. Both experimental and theoretical investigations of charge imbalance have focused mostly on temperatures near the critical temperature $T_\mathrm{c}$ of the superconductor. In this regime, charge imbalance is easily accessible to experiments basically due to the divergence of the signal towards $T_\mathrm{c}$, and excellent theoretical approximations are available.\cite{clarke1972,tinkham1972,langenberg} Recently, the investigation of non-local transport properties of superconductors has gained new impetus from two separate but loosely related fields. One is the investigation of spin-dependent transport,\cite{johnson1994,*valenzuela2006,*poli2008,*luo2009} and in particular spin diffusion and accumulation in the context of spintronics. The second is the investigation of coherent non-local effects such as crossed Andreev reflection,\cite{byers1995,*deutscher2000,*beckmann2004,*beckmann2007,*russo2005,*cadden2006,*cadden2007,*cadden2009,*asulin2006,*kleine2009,*hofstetter2009,*herrmann2010,*huebler2010} which might be useful for quantum information processing. In the diffusive quasi-onedimensional structures typically used for such experiments, the magnitude of the signals due to these phenomena scale with the normal-state resistance of the superconductor over a characteristic length scale, which is given by the charge-imbalance relaxation length $\lambda_{Q^*}\sim 10~\mathrm{\mu m}$,\cite{langenberg} the spin-diffusion length $\lambda_\mathrm{sf}\sim 1~\mathrm{\mu m}$,\cite{johnson1985,valet1993,jedema2002,jedema2003} and the coherence length $\xi\sim 0.1~\mathrm{\mu m}$,\cite{golubev2009} respectively. Consequently, the signals due to charge imbalance are often the largest, and make an unambiguous identification of the other phenomena difficult. In particular for the investigation of crossed Andreev reflection, experiments far below $T_\mathrm{c}$ are necessary. Despite the vast amount of experimental and theoretical literature on charge imbalance, surprisingly little is known about the subject at very low temperatures.\cite{yagi2006} This is probably due to the fact that no simple theoretical models are available for this regime, and that the interpretation of the widely-used non-local resistance measurement scheme becomes increasingly difficult as temperature is lowered.

In this article, we report on a detailed investigation of non-local {\em conductance} rather than {\em resistance} in superconductor/normal-metal hybrid structures at temperatures $T\ll T_\mathrm{c}$. The samples consist of a quasi-onedimensional superconducting wire, with several normal-metal tunnel junctions attached to it. From these experiments, we deduce the charge-imbalance relaxation length $\lambda_{Q^*}$ as a function of bias voltage, temperature and magnetic field. The results allow a detailed comparison to theoretical predictions for different charge-imbalance relaxation mechanisms, and and an assessment of the possible impact on the interpretation of experiments on crossed Andreev reflection and spin diffusion. 

\begin{figure}
\includegraphics[width=\columnwidth]{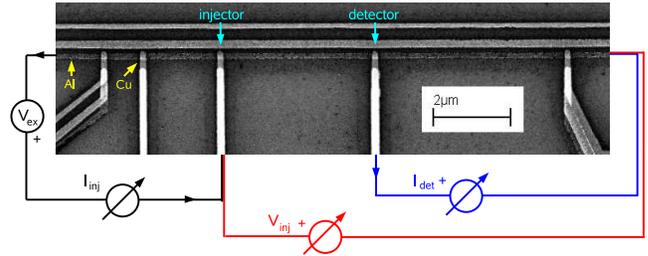}
\caption{\label{fig_sem}(Color online) SEM image of sample A illustrating the experimental scheme. 
Five copper (Cu) fingers are connected by tunnel contacts to an aluminum bar (Al). For one of the possible injector-detector pairs the bias and measurement scheme used for charge imbalance detection is shown.}
\end{figure}

\section{Theory}\label{sec_theory}

Charge imbalance (CI) can be described using two different theoretical frameworks, the quasiparticle, or two-fluid, approach,\cite{clarke1972,tinkham1972,tinkham1972b,pethick1979} and quasiclassical Green's functions.\cite{schmid1975} The relation of these two approaches has been discussed extensively in the literature.\cite{clarke1979,entin1979} We simply note here that the Green's function method is more general, and can be applied, e.g., to inhomogeneous superconducting states and situations with strong pair breaking. We will nevertheless use the quasiparticle approach, due to its conceptual (and computational) simplicity, and discuss its shortcomings where necessary.

We consider a quasi-onedimensional superconductor of length $L$ along the $x$ axis, with several normal-metal electrodes attached via tunnel junctions. These electrodes will serve both to inject nonequilibrium quasiparticles into the superconductor, and to detect them. The quasiparticle energies $E$ are given by
\begin{equation}
E=\sqrt{\epsilon^2+\Delta^2},
\end{equation}
where $\epsilon$ is the normal-state electron energy relative to the Fermi energy, and $\Delta$ is the pair potential. The normalized quasiparticle density of states is
\begin{equation}
n(E)=\Theta(E-\Delta)\frac{E}{|\epsilon|},
\end{equation}
where $\Theta$ is the Heaviside function. Charge imbalance is defined as
\begin{equation}
Q^* = 2N_0\int_{-\infty}^{\infty}q(\epsilon)f(\epsilon)d\epsilon,\label{equ_qstar}
\end{equation}
where $N_0$ is the density of states per spin at the Fermi energy in the normal state, $q(\epsilon)=\epsilon/E$ is the effective quasiparticle charge in units of the elementary charge $e$, and $f(\epsilon)$ is the quasiparticle distribution function. By using $\epsilon$ rather than $E$ as independent variable, we keep track of the electron-like ($\epsilon>0$) and hole-like ($\epsilon<0$) branch of the quasiparticle spectrum. In thermal equilibrium, $f(\epsilon)$ is given by the Fermi distribution $f_\mathrm{T}(E)$. 

It is apparent from (\ref{equ_qstar}) that $Q^*$ is non-zero only if the populations of the electron- and hole-like branches are unequal, i.e., $f(\epsilon)-f(-\epsilon)\not\equiv 0$. Besides this charge-mode (transverse) non-equilibrium, there is also an energy-mode (longitudinal) non-equilibrium characterized by $f(\epsilon)+f(-\epsilon)-2f_\mathrm{T}(E)\not\equiv 0$. As will be shown below, the latter enters transport properties only indirectly via the self-consistency equation for the pair potential,
\begin{equation}
1+\mathcal{V}\int_{-\infty}^{\infty}\frac{1-2f(\epsilon)}{2\sqrt{\epsilon^2+\Delta^2}}d\epsilon=0,\label{equ_sc}
\end{equation}
where $\mathcal{V}$ is the pairing interaction.

The electric current through a tunnel junction between a normal metal held at bias voltage $V$ and a nonequilibrium superconductor is given by the sum of the usual "Giaever" tunnel current $I_\mathrm{T}(V)$,\cite{giaever1960} and a bias-independent extra current $I_{Q^*}$ due to CI,\cite{tinkham1972b}
\begin{equation}
I(V)=I_\mathrm{T}(V)+I_{Q^*}.\label{equ_current}
\end{equation}
Here,
\begin{equation}
I_\mathrm{T}(V) =\frac{G_\mathrm{N}}{e} \int\limits_{0}^\infty n(E) \left( f_\mathrm{T}(E-eV)-f_\mathrm{T}(E+eV) \right)dE \label{equ_itunnel}
\end{equation}
where $G_\mathrm{N}$ is the normal-state tunnel conductance, and the Fermi functions describe the electron occupation in the normal metal. The excess current is given by
\begin{equation}
I_{Q^*} =\frac{G_\mathrm{N}Q^*}{2eN_0}.\label{equ_iq}
\end{equation}
As mentioned above, the current $I_{Q^*}$ depends on the nonequilibrium distribution $f(\epsilon)$ only via the charge mode, whereas $I_\mathrm{T}(V)$ indirectly depends on the energy mode via the gap equation (\ref{equ_sc}).

In the vicinity of the critical temperature $T_\mathrm{c}$ of the superconductor, the deviation of the distribution function $f(\epsilon)$ from equilibrium is small and can be approximated by a Fermi function with a shift $\Delta\mu$ of the chemical potential of the quasiparticles relative to the chemical potential of the Cooper pairs.\cite{pethick1979} The most convenient measurement technique in this situation is the widely used non-local voltage detection scheme, in which the voltage between a normal-metal detector junction and the superconductor is measured. The voltage adjusts such that $I(V_\mathrm{det})=0$, i.e., the current $I_\mathrm{Q^*}$ is cancelled by the backflow $I_\mathrm{T}(V_\mathrm{det})$. In the regime where the chemical-potential model applies, this voltage is given by $eV_\mathrm{det}=\Delta\mu$, i.e., it directly measures the single parameter that characterizes CI. Voltage detection, however, is less useful in the low-temperature regime that we are interested in for two reasons. First, at low temperature the chemical-potential model breaks down, and the detector voltage has no longer a simple physical meaning. Second, the non-linearity and temperature dependence of the tunnel current $I_\mathrm{T}$ distort the measured signal. This has already been noted in the earliest experiment on charge imbalance,\cite{clarke1972} where the raw data at low temperature had to be corrected by the temperature dependence of the detector junction for comparison with theory. To avoid these shortcomings, we measure the detector current $I_\mathrm{det}(V_\mathrm{det}=0)=I_\mathrm{Q^*}$. Here, the only detector property that enters is the (constant) normal-state conductance, and therefore $I_\mathrm{det}$ is a direct measure of $Q^*$ even for arbitrary non-equilibrium distributions. Experimentally, we measure the differential non-local conductance, and we will now derive the expression used to evaluate our results.

The distribution function $f(\epsilon)$ is driven out of equilibrium by tunnel injection into a fixed volume $\Omega$ of the superconductor at a rate given by\cite{tinkham1972b,chi1980}
\begin{align}
\frac{\partial f}{\partial t}=\gamma_\mathrm{tun}=&\frac{1}{\tau_\mathrm{tun}}\bigg\{ 
\frac{1}{2}\left(1+\frac{\epsilon}{E}\right)\left(f_\mathrm{T}(E-eV)-f(\epsilon)\right)\nonumber\\*
& -\frac{1}{2}\left(1-\frac{\epsilon}{E}\right)\left(f(\epsilon)-f_\mathrm{T}(E+eV)\right)\bigg\},\label{equ_gamma_inj}
\end{align}
where $\tau_\mathrm{tun}^{-1}=G_\mathrm{N}/2N_0\Omega e^2$. Here, we use the full non-equilibrium distribution $f(\epsilon)$ in the superconductor, rather than the thermal distribution used in Ref. \onlinecite{chi1980}. However, we retain the assumption that the normal-metal electrode is at thermal equilibrium, described by the shifted Fermi functions $f_\mathrm{T}(E\pm eV)$. It is customary to define an injection efficiency $F^*=e\Omega\dot{Q}^*/I$. Since we will be interested in the differential conductance, we use the spectral quantity $f^*(E)=\epsilon^2/E^2$ rather than the usual integral definition. The expression for $f^*$ neglects the non-equilibrium contributions in (\ref{equ_current}) and (\ref{equ_gamma_inj}), which are unimportant for an injector junction biased at $|eV|\gtrsim \Delta$. On the other hand, a detector junction held at $V=0$ leads to charge imbalance relaxation at a rate $\tau_\mathrm{tun}^{-1}$, and care must be taken to ensure that this rate is negligible compared to bulk relaxation mechanisms for non-invasive detection.\cite{lemberger1984} Once injected, non-equilibrium quasiparticles diffuse along the wire with an energy-dependent diffusion constant $D(E)=v_\mathrm{g}D_\mathrm{N}$,\cite{entin1979,ullom1998} where $D_\mathrm{N}$ is the normal-state diffusion constant, and $v_\mathrm{g}=|\epsilon|/E$ is the normalized group velocity of quasiparticles. Concomitantly, the non-equlibrium distribution relaxes due to different mechanisms, including inelastic electron-phonon scattering,\cite{tinkham1972,tinkham1972b,kaplan1976} elastic impurity scattering in the presence of gap anisotropy,\cite{tinkham1972b,chi1979} and magnetic pair breaking.\cite{schmid1975,lemberger1981,lemberger1981b} Charge-imbalance relaxes over a characteristic time $\tau_{Q^*}=Q^*/\dot{Q}^*$, which we will also assume to be energy-dependent, leading to an exponential relaxation on the length scale $\lambda_{Q^*}=\sqrt{D\tau_{Q*}}$. The steady-state distribution is achieved when injection and relaxation rates are equal. Assuming that the effective injection volume is given by a wire section of length $2\lambda_{Q^*}$, we find that the non-local conductance due to charge imbalance is given by
\begin{equation}\label{equ_gnl}
g_\mathrm{nl} =\frac{dI_\mathrm{det}}{dV_\mathrm{inj}} = g^*G_\mathrm{inj}G_\mathrm{det}\frac{\rho_\mathrm{N}\lambda_{Q^*}}{2A}\exp\left(-\frac{d}{\lambda_{Q^*}}\right),
\end{equation}
where $G_\mathrm{inj}$ and $G_\mathrm{det}$ are the normal-state conductances of the injector and detector junctions, $\rho_\mathrm{N}$ is the normal-state resistivity of the superconductor, and $A$ is the cross-section of the wire. The factor $g^*$ accounts for thermal smearing, injection efficiency, etc., and is of order unity. At $T=0$, and neglecting energy relaxation, $g^*=n(E)f^*/v_\mathrm{g}=\Theta(eV_\mathrm{inj}-\Delta)$.  Equation (\ref{equ_gnl}) will form the basis for our data analysis.

\section{Experiment}\label{sec_experiment}

Fig.~\ref{fig_sem} shows the relevant part of one of the three samples discussed, together with a scheme of the measurement setup.  All samples are fabricated by e-beam lithography and shadow evaporation techniques. In the following, the processing details are explained with the help of the sample parameters listed in Table \ref{tab_prop}.

\begin{table}
\caption{\label{tab_prop}Characteristic parameters of the three samples A-C.
Aluminum film thickness $t_\mathrm{Al}$, width $w_\mathrm{Al}$, and normal state resistivity $\rho_\mathrm{Al}$ at $T=4.2~\mathrm{K}$, range of contact distances $d$ and contact conductances $G$, critical temperature $T_\mathrm{c}$, critical field $B_\mathrm{c}$ and energy gap $\Delta_0$.}
\begin{ruledtabular}
\begin{tabular}{lcccccccc}
  & $t_\mathrm{Al}$ & $w_\mathrm{Al}$ & $\rho_\mathrm{Al}$   & $d$ & $G$     & $T_\mathrm{c}$ & $B_\mathrm{c}$ & $\Delta_0$ \\ 
       & (nm) & (nm) & ($\mathrm{\mu \Omega cm}$) & ($\mu$m) & ($\mu$S) & (K) & (T) & ($\mathrm{\mu eV}$) \\ \hline
A      & 30   & 140   & 4.9  & 1-12      & 230-270  & 1.39             & 0.53  & 208 \\
B      & 30   & 190   & 4.9  & 0.2-9.7   & 260-350  & 1.38             & 0.58  & 218   \\
C      & 12.5 & 140   & 11.1 & 0.5-6.5   & 370-490  & 1.5              & 1.73  & 225 \\
\end{tabular}
\end{ruledtabular}
\end{table}

First, a copper film with thickness $t_\mathrm{Cu1}=25-30~\mathrm{nm}$ is evaporated onto a thermally oxidized silicon substrate. This first layer will form Ohmic interconnections to the subsequent layers. In a second evaporation step, the superconductor, an aluminum bar of thickness $t_\mathrm{Al}$ and width $w_\mathrm{Al}$, is deposited under a different angle, shifting the design to create intended overlaps only. In order to provide the formation of an insulating layer on the top, the aluminum is then oxidized {\em in situ} by applying the equivalent of $1~$Pa of oxygen for $10~$min. In the final evaporation step, a second layer of copper ($t_\mathrm{Cu2}$ = 30~nm) is deposited under a third angle forming five tunnel contacts with the aluminum. The contact distances between neighbouring copper fingers presented in sample A are about 1~$\mu$m, 2~$\mu$m, 4~$\mu$m and 5~$\mu$m from left to right, respectively.

For the transport experiment the samples are mounted into a shielded box thermally anchored to the mixing chamber of a dilution refrigerator. The measurement lines are fed through a series of filters to eliminate rf and microwave radiation from the shielded box.
A voltage $V_\mathrm{ex}$ consisting of a dc bias and a low-frequency ac excitation is applied to the injector contact, and the ac part of the resulting current $I_\mathrm{inj}$ is measured with a lock-in technique. Simultaneously, the ac current $I_\mathrm{det}$ is measured through the second contact, the detector. The local and non-local differential conductances $g_\mathrm{inj}=dI_\mathrm{inj}/dV_\mathrm{inj}$ and $g_\mathrm{nl}=dI_\mathrm{det}/dV_\mathrm{inj}$ are extracted from the ac signals. 
Voltage and current polarities are indicated in Fig.~\ref{fig_sem} by plus signs and arrows, respectively.
All contacts are measured in a three-point configuration with a series resistance of about 90~$\Omega$ coming from the measurement line.

\section{Results}\label{sec_results}

\subsection{Contact and film characterization}

\begin{figure}
\includegraphics[width=\columnwidth]{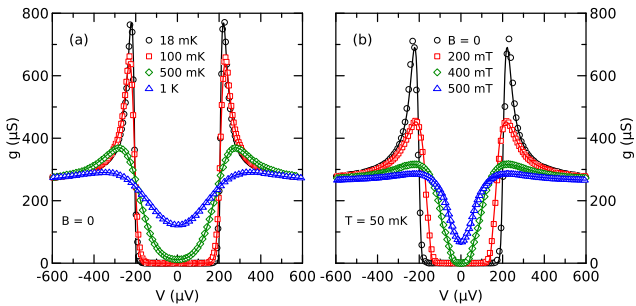}
\caption{\label{fig_local}(Color online)
Local differential conductance $g=dI/dV$ of one contact of sample B as a function of bias voltage $V$ for (a) different temperatures $T$ and (b) different applied magnetic fields $B$. Symbols represent measured data, lines are fits to the model described in the text.}
\end{figure}

To characterize the tunnel contacts and the properties of the superconducting film, we first discuss the tunnel spectra of the individual junctions. Fig.~\ref{fig_local} shows the local differential conductance $g_\mathrm{inj}$ as a function of bias $V_\mathrm{inj}$ for one of the tunnel junctions of sample B. Panel (a) represents the temperature dependence at zero magnetic field, whereas panel (b) depicts the variation with magnetic field applied in the plane of the substrate along the Cu wires for constant temperature $T = 50~\mathrm{mK}$. At lowest temperature and zero field, the differential conductance is completely suppressed at low bias, with sharp peaks at the energy gap, showing the high quality of the oxide tunnel barrier. Upon increasing the temperature or the magnetic field, the features are broadened and the gap is reduced. Since we are particularly interested in the dependence on magnetic field, we have used a slightly more elaborate model than (\ref{equ_itunnel}) to fit the data. In the presence of a magnetic field, the spin-resolved density of states in the superconductor can be described by\cite{maki1964b}
\begin{equation}
n_{\pm}(E)=\frac{1}{2}\mathrm{Re}\left(\frac{u_\pm}{\sqrt{u_\pm^2-1}}\right),
\end{equation}
where the complex quantities $u_\pm$ have to be determined from the implicit equation
\begin{equation}
\frac{E\mp\mu_\mathrm{B}B}{\Delta}=u_\pm\left(1-\frac{\Gamma}{\Delta}\frac{1}{\sqrt{1-u_\pm^2}}\right)
+b_\mathrm{so}\left(\frac{u_\pm-u_\mp}{\sqrt{1-u_\mp^2}}\right).\nonumber
\end{equation}
Here, $\mu_\mathrm{B}$ is the Bohr magneton, $\Gamma$ is the pair-breaking parameter, $b_\mathrm{so}=\hbar/3\tau_\mathrm{so}\Delta$ measures the spin-orbit scattering strength, and we have dropped a small higher-order term. The fits in Fig.~\ref{fig_local} were obtained by replacing the BCS density of states $n(E)$ by $n_+(E)+n_-(E)$ in (\ref{equ_itunnel}). During fitting, $T$ and $B$ were taken from the experiment, and the remaining parameters were varied. Including the Zeeman splitting was found to be necessary for the data at higher fields. The small spin-orbit term gave minor improvements of the fits, but could not be determined precisely. We simply chose a suitable value $b_\mathrm{so}\sim O(0.01)$ at high field, and kept it fixed for all other fits. Similar values can be found in the literature.\cite{meservey1975} The quality of the fits was excellent for samples A and B, as shown for the latter in Fig.~\ref{fig_local}. For the thin-film sample C, the quality of the fits was rather poor, and no reliable parameters could be obtained.

\begin{figure}
\includegraphics[width=\columnwidth]{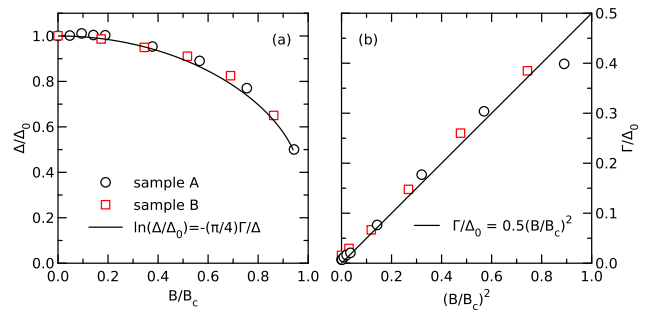}
\caption{\label{fig_gamma}(Color online)
(a) Pair potential $\Delta$ as a function of the applied magnetic field $B$. $\Delta$ is normalized to its zero-field value $\Delta_0$, and $B$ is normalized to the critical field $B_\mathrm{c}$. 
(b) Normalized pair-breaking parameter $\Gamma/\Delta_0$ as a function of the $(B/B_\mathrm{c})^2$. 
The lines are predictions from pair-breaking theory.}
\end{figure}

For samples A and B, where the fits were reliable, we show the pair-breaking parameter $\Gamma$ as well as the pair potential $\Delta$ as a function of the magnetic field in Fig.~\ref{fig_gamma}. In panel (a), the normalized pair potential $\Delta/\Delta_0$ is displayed as a function of $B/B_\mathrm{c}$, together with the expectation $\ln(\Delta/\Delta_0)=-(\pi/4)(\Gamma/\Delta)$.\cite{abrikosov1961} The critical pair-breaking strength for the suppression of superconductivity is given by $2\Gamma=\Delta_0$,\cite{abrikosov1961} and together with $\Gamma\propto B^2$ for a thin film in parallel magnetic field,\cite{maki1964a} we can rewrite $\Gamma$ as
\begin{equation}
\frac{\Gamma}{\Delta_0}=\frac{1}{2}\left(\frac{B}{B_\mathrm{c}}\right)^2.\label{equ_gammanrm}
\end{equation}
Fig.~\ref{fig_gamma}(b) shows $\Gamma/\Delta_0$ as a function of $(B/B_\mathrm{c})^2$. The solid line is the theoretical expectation (\ref{equ_gammanrm}). As can be seen, samples A and B can be described perfectly well by standard pair-breaking theory, and we will assume (\ref{equ_gammanrm}) to hold also for sample C later on.

\subsection{Energy-mode non-equilibrium}

\begin{figure}
\includegraphics[width=\columnwidth]{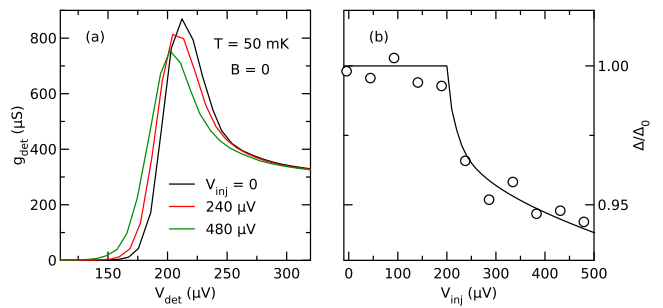}
\caption{\label{fig_inj}(Color online) 
a) Differential conductance $g_\mathrm{det}$ of the left-most contact of sample A as a function of bias voltage $V_\mathrm{det}$, with additional voltage bias $V_\mathrm{inj}$ applied to the neighbour contact at $1~\mathrm{\mu m}$ distance. b) Normalized pair potential $\Delta/\Delta_0$ as a function of injector bias $V_\mathrm{inj}$. The line is a guide to the eye.}
\end{figure}

While we are mostly interested in charge imbalance, we have also investigated the impact of energy-mode non-equilibrium in our samples. To this effect, we monitor the differential conductance $g_\mathrm{det}$ of a detector junction while non-equilibrium quasiparticles are injected through a second nearby injector contact. Fig.~\ref{fig_inj}(a) shows the differential conductance $g_\mathrm{det}$ of the left-most contact of sample A (see Fig.~\ref{fig_sem}) as a function of the local bias voltage $V_\mathrm{det}$ for different injector bias $V_\mathrm{inj}$ applied to the neighbouring contact at a distance of $1~\mathrm{\mu m}$. We focus here only on the bias region of the gap features. As the injector bias is increased, the density-of-states peak in the conductance shifts to lower bias, and broadens slightly. The increased broadening of the gap features might signify an increased temperature of the normal-metal side of the junction due to the injection of non-equlibrium quasiparticles tunneling out of the superconductor.\cite{takane2007} An alternative explanation would be an increased lifetime broadening of the density of states of the superconductor due to scattering of non-equilibrium quasiparticles. From our data, we can not make a clear decision between these scenarios. The evolution of the pair potential $\Delta$ as a function of injector bias is plotted in panel (b), normalized to its value $\Delta_0$ at $V_\mathrm{inj}=0$. No significant change is observed for $eV_\mathrm{inj}<\Delta_0$. As soon as $eV_\mathrm{inj}$ exceeds $\Delta_0\approx 200~\mathrm{\mu eV}$, the energy gap drops quickly by a few percent, and continues to decrease more slowly for higher bias. The gap reduction can be understood from the inspection of the self-consistency equation (\ref{equ_sc}). For injector voltages in the vicinity of $\Delta$, a large number of quasiparticles are injected due to the divergence of the density of states. In addition, these quasiparticles are very efficient in reducing the gap due to the energy denominator in the integral. Therefore, the initial decrease is steep, and then becomes more shallow as the density of states flattens, and the denominator increases.

\subsection{Charge-mode non-equilibrium}

\begin{figure}
\includegraphics[width=\columnwidth]{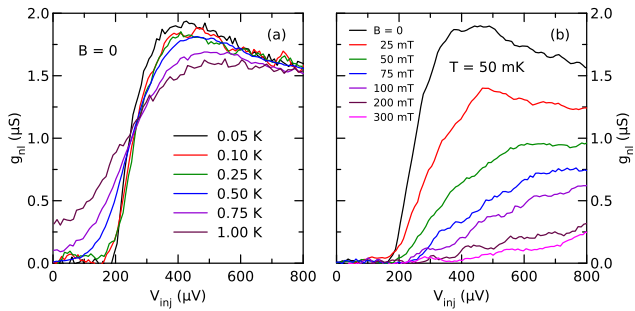}
\caption{\label{fig_nonlocal}(Color online) 
Non-local differential conductance $g_\mathrm{nl}=dI_\mathrm{det}/dV_\mathrm{inj}$ for an injector/detector pair of sample A as a function of bias voltage $V_\mathrm{inj}$ (a) for different temperatures $T$ and (b) for different applied magnetic fields $B$.}
\end{figure}

Figure \ref{fig_nonlocal} displays the non-local conductance $g_\mathrm{nl}$ as a function of the injector bias $V_\mathrm{inj}$ for one injector/detector-pair of sample A. Panel (a) shows data for different temperatures $T$ without an applied magnetic field. At $T=50~\mathrm{mK}$, the non-local conductance is zero within the experimental resolution for bias voltages below $\Delta/e \approx 200~\mathrm{\mu V}$. Above the energy gap, the signal increases continuously from zero with a finite initial slope up to a broad maximum at $V_\mathrm{inj} \approx 450~\mathrm{\mu V}$ before it decreases slowly again. With increasing temperature, the signal smears out around the gap, whereas the value at high bias remains unchanged. Panel (b) shows the impact of a magnetic field $B$ applied in the substrate plane along the direction of the copper wires. In contrast to temperature, the signal depends strongly on the magnetic field. The initial slope decreases, and the maximum decreases and shifts to higher bias until it is no longer observable within our bias range for $B\geq 100~\mathrm{mT}$.

\begin{figure}
\includegraphics[width=\columnwidth]{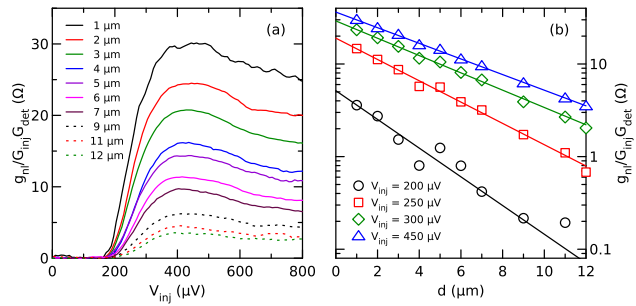}
\caption{\label{fig_dist}(Color online) (a) Normalized nonlocal differential conductance $g_\mathrm{nl}/G_\mathrm{inj}G_\mathrm{det}$ as a function of injector bias voltage $V_\mathrm{inj}$ for different contact distances $d$. (b) Semi-logarithmic plot of $g_\mathrm{nl}/G_\mathrm{inj}G_\mathrm{det}$ as a function of contact distance $d$ for different injector bias $V_\mathrm{inj}$. The solid lines are fits to (\ref{equ_gnl}). 
}
\end{figure}

Figure \ref{fig_dist}(a) shows the non-local differential conductance for several injector/detector contact pairs of sample A at $T=50~\mathrm{mK}$ and $B=0$. Since $g_\mathrm{nl}\propto G_\mathrm{inj}G_\mathrm{det}$, we have normalized the data accordingly to exclude the impact of small variations of the junction conductances. The overall signal magnitude decreases with increasing contact distance while the shape remains unchanged. Panel (b) shows the normalized non-local conductance as a function of contact distance $d$ for different injector bias on a semi-logarithmic scale. The solid lines are fits to the exponential decay predicted by (\ref{equ_gnl}). The quality of the fits is generally good, except for the very small signals at lowest bias. From the fits, the relaxation length $\lambda_{Q^*}$ and the amplitude
\begin{equation}\label{equ_amplitude}
a=g^*\frac{\rho_\mathrm{N}\lambda_{Q^*}}{2A}
\end{equation} 
can be extracted.

\begin{figure}
\includegraphics[width=\columnwidth]{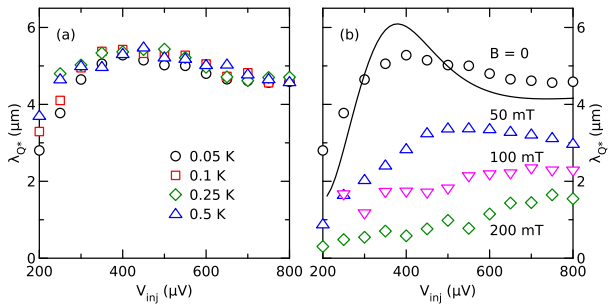}
\caption{\label{fig_lambda}(Color online)
Charge imbalance relaxation length $\lambda_{Q^*}$ as a function of injector bias voltage $V_\mathrm{inj}$ for (a) different temperatures $T$ and (b) different applied magnetic fields $B$. The line is the result of a numerical simulation described in section \ref{sec_discussion}.}
\end{figure}

The charge imbalance relaxation length $\lambda_{Q^*}$ extracted from these fits is shown in Fig.~\ref{fig_lambda} as a function of injector bias for different temperatures $T$ (a) and magnetic fields $B$ (b). The data resemble those of the non-local conductance shown in Fig.~\ref{fig_nonlocal}. This is not surprising, since the signal amplitude $a$ is itself proportional to $\lambda_{Q^*}$. A noticeable difference is that $\lambda_{Q^*}$ is nearly independent of temperature. This indicates that the temperature dependence seen in Fig.~\ref{fig_nonlocal}(a) is mostly due to thermal broadening of the distribution in the injector contact rather than a change in relaxation rates. In contrast, the suppression of the non-local conductance upon increasing the magnetic field is reflected in the pronounced field dependence of $\lambda_{Q^*}$, indicating an increase of the relaxation rate.

\begin{table}
\caption{\label{tab_results}Maximum values of the relaxation length $\lambda_{Q^*}$, the relaxation time $\tau_{Q^*}$,
and the ratio $\tau_{Q^*}/\tau_\mathrm{tun}$ for all three samples.}
\begin{ruledtabular}
\begin{tabular}{lccc}
sample & $\lambda_{Q^*}$ & $\tau_{Q^*}$ & $\tau_{Q^*}/\tau_\mathrm{tun}$ \\ 
       & ($\mu$m) & (ns) &   \\ \hline
A      & 5.2   & 7.8   & 0.01  \\
B      & 4.3   & 5.2   & 0.01  \\
C      & 3.0   & 5.9   & 0.05  \\
\end{tabular}
\end{ruledtabular}
\end{table}

Similar results to those presented in Figs.~\ref{fig_nonlocal}-\ref{fig_lambda} were obtained for all three samples. From $\lambda_{Q^*}$, we can calculate $\tau_{Q^*}=\lambda_{Q^*}^2/D_\mathrm{N}v_\mathrm{g}$. The maximum values of $\lambda_{Q^*}$ and $\tau_{Q^*}$ obtained at lowest temperature and zero magnetic field are listed in table \ref{tab_results}, along with the maximum of the ratio $\tau_{Q^*}/\tau_\mathrm{tun}$. We find $\tau_{Q^*}/\tau_\mathrm{tun}\ll 1$ for all samples, confirming that our detector junctions are non-invasive.

\begin{figure}
\includegraphics[width=\columnwidth]{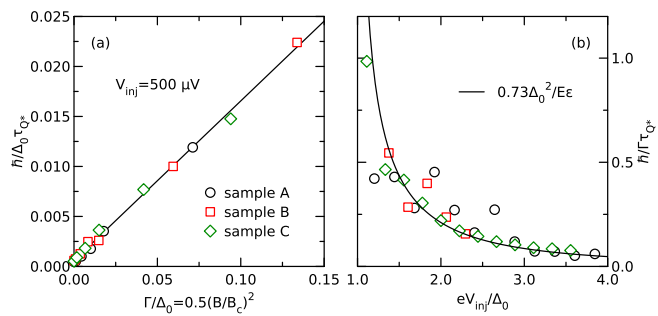}
\caption{\label{fig_tau}(Color online)
(a) Normalized charge-imbalance relaxation rate $\hbar/\Delta_0\tau_{Q^*}$ as a function of the normalized pair-breaking parameter $\Gamma/\Delta_0$. The line is a guide to the eye.
(b) Normalized charge-imbalance relaxation rate $\hbar/\Gamma\tau_{Q^*}$ as a function of normalized injector bias $eV_\mathrm{inj}/\Delta_0$. The line is a joint fit to the data of all three samples.}
\end{figure}

We will now focus in more detail on the suppression of charge imbalance as a function of magnetic field. Figure \ref{fig_tau}(a) shows the normalized charge-imbalance relaxation rate $\hbar/\Delta_0\tau_{Q^*}$ as a function of the normalized pair-breaking parameter $\Gamma/\Delta_0$ for fixed injector bias. Here we have made use of equation (\ref{equ_gammanrm}) to calculate $\Gamma$ from $B$ for all three samples. The data from all samples fall onto a single line, and the relaxation rate at zero field is negligible on the scale of the plot. We note that in the magnetic-field range of the plot ($B\lesssim 0.5B_\mathrm{c}$) the spectral properties of the superconductor remain almost unchanged. The reduction of $\Delta$, for example, is less than $10\%$ in this range. The relaxation rate due to elastic pair-breaking perturbations such as magnetic impurities,\cite{schmid1975,artemenko1978,lemberger1981} supercurrent\cite{lemberger1981b} and applied magnetic field\cite{takane2008} has been calculated both within the quasiparticle description used by us, and from quasi-classical Green's functions. We note that by convention, rates from the Green's function formalism differ from those of the quasiparticle description by the factor $f^*$.\cite{clarke1979} When properly adjusted, the rate is predicted to be
\begin{equation}
\frac{1}{\tau_\mathrm{Q^*}}= \alpha\frac{\Gamma}{\hbar}\frac{\Delta^2}{E\epsilon},\label{equ_taub}
\end{equation}
where $\alpha$ is a numerical prefactor of order unity which we will use as a fit parameter. From linear fits of the data in panel (a) we can extract $\hbar/\Gamma\tau_{Q^*}$. The result extracted from such fits is plotted in panel (b) as a function of normalized injector bias $eV_\mathrm{inj}/\Delta_0$ for all three samples. The solid line is a joint fit of all data to (\ref{equ_taub}) (where we have set $E=eV_\mathrm{inj}$ and $\Delta=\Delta_0$), with $\alpha=0.73$.

\begin{figure}
\includegraphics[width=\columnwidth]{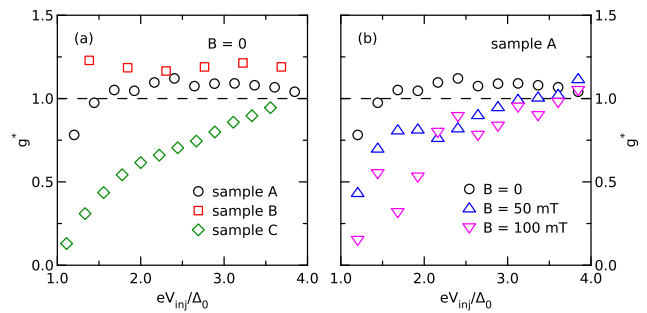}
\caption{\label{fig_gstar}(Color online) $g^*$ as a function of normalized injector bias $eV_\mathrm{inj}/\Delta_0$ (a) for all samples at $B=0$ and (b) for sample A at different magnetic fields $B$.}
\end{figure}

From the signal amplitude $a$ extracted from the fits in Fig.~\ref{fig_dist}, we can calculate the prefactor $g^*$ using (\ref{equ_amplitude}) together with known sample parameters and $\lambda_{Q^*}$ extracted from the same fits. The results are plotted in Fig.~\ref{fig_gstar} as a function of normalized injector bias (a) for all samples at $B=0$ and (b) for sample A at different magnetic fields. At $B=0$, $g^*$ follows the expectation $g^*\approx \Theta(eV_\mathrm{inj}-\Delta)$ for samples A and B, whereas it deviates at low bias both for sample C, and in the presence of a magnetic field. Since $g^*$ depends on a combination of several spectral properties of the superconductor, we can not identify the precise cause of these deviations. For sample C, the enhanced energy-mode non-equilibrium due to the reduced film thickness may play a role. However, in all cases $g^*\approx 1$ at high bias, which justifies our choice of using a wire section of length $2\lambda_{Q^*}$ as injection volume. 

\section{Discussion}\label{sec_discussion}

The bias dependence of the non-local conductance can be understood as follows: Charge-imbalance relaxation takes place mostly at energies close to $\Delta$, since the coherence factors for scattering between the electron- and hole-like branches, both elastic and inelastic, vanish at higher energies. In addition, in the low-temperature regime the inelastic contribution is expected to be negligible.\cite{takane2006} The elastic relaxation rate, both due to gap anisotropy and magnetic pair breaking, diverges for $E\rightarrow \Delta$, and quickly drops at higher energies. Consequently, charge imbalance rises continuously from zero as the bias is increased above $\Delta$, and in this regime relaxation is mainly due to direct scattering between the branches. As the bias is increased further, the direct relaxation rate decreases. On the other hand, energy relaxation due to inelastic scattering becomes important, and charge relaxation turns into a two-stage process.\cite{tinkham1972} Quasiparticles are first cooled to the vicinity of $\Delta$ by inelastic scattering, and are then scattered elastically between branches. If we assume that inelastic scattering is described by electron-phonon scattering in the Debye approximation, the inelastic rate quickly increases as more phonons become available at higher energy. Consequently, the non-local differential conductance begins to drop again after its initial increase. The bias at which the maximum appears scales roughly with the transition between the direct and two-stage relaxation regimes. The temperature dependence, or lack thereof, can be understood easily from this picture. Elastic relaxation depends only weakly on temperature via $\Delta$. For inelastic relaxation, at low temperatures the Bose factors for emission and absorption of phonons become $\approx 1$ and $\approx 0$, respectively. Relaxation is then dominated by phonon emission at a rate which only depends on the bias-dependent energy of the quasiparticles, but not on temperature. Consequently, the relaxation rate, and $\lambda_{Q^*}$, become practically independent of temperature over the entire bias range, as observed in Fig.~\ref{fig_lambda}(a). We have identified the quasiparticle energy with the bias voltage throughout our data analysis in section \ref{sec_results}. This has to be taken with some caution due to the presence of inelastic energy relaxation. Our approximation mainly affects the calculation of $\tau_{Q^*}=\lambda_{Q^*}^2/D_\mathrm{N}v_\mathrm{g}$, where $v_\mathrm{g}$ should actually be an average over energy. However, since inelastic relaxation is important mostly at higher energies, where $v_\mathrm{g}\approx 1$, we assume that the error is small. This is corroborated by the observation that the position of the maximum in $g_\mathrm{nl}$ does not depend much on contact distance, as seen in Fig.~\ref{fig_lambda}(a). We would nevertheless like to stress that $\lambda_{Q^*}$ and $\tau_{Q^*}$ are not really spectral quantities, but depend in detail on the non-equilibrium distribution and bias conditions.

To make a quantitative connection to microscopic theory, we have performed numerical simulations of the quasiparticle Boltzmann equation, basically following Ref.~\onlinecite{chi1979}. We have used the onedimensional form of the Boltzmann equation\cite{chi1979,entin1979}
\begin{equation}
\frac{\partial f}{\partial t}-v_\mathrm{g}D_\mathrm{N}\frac{\partial^2 f}{\partial x^2}=\gamma_\mathrm{tun}-\gamma_\mathrm{el}-\gamma_\mathrm{in}.\label{equ_boltzmann}
\end{equation}
The elastic and inelastic relaxation rates $\gamma_\mathrm{el}$ and $\gamma_\mathrm{in}$ are given by equations (2.16) and (2.17) of Ref.~\onlinecite{chi1979}, respectively. The injection rate is given by (\ref{equ_gamma_inj}). Steady-state solutions $f(\epsilon,x)$ of the Boltzmann equation (\ref{equ_boltzmann}) were obtained by numerical iteration on a discretized grid. Both injector and detector junction were included on an equal footing, with injection rates and currents given by (\ref{equ_gamma_inj}) and (\ref{equ_current}). Here, the injection volume $\Omega$ is given by the grid point size, and the spreading of charge imbalance over the length scale $\lambda_{Q^*}$ is included microscopically in the diffusion term. The granularity of the grid was chosen sufficiently small ($\Delta\epsilon=20~\mathrm{\mu eV}$, $\Delta x=500~\mathrm{nm}$) not to affect the results. Parameters such as wire geometry, diffusion constant, contact conductances $G_\mathrm{inj}$ and $G_\mathrm{det}$, etc., were taken directly from the experiment, leaving only the characteristic electron-phonon scattering time $\tau_0$ and the average gap anisotropy $\left<a^2\right>_0$ as free parameters. The results of the simulation are shown as a solid line in Fig.~\ref{fig_lambda}(b), where we have inserted the typical values $\tau_0=100~\mathrm{ns}$ and $\left<a^2\right>_0=0.03$ from Ref.~\onlinecite{chi1979}. The overall magnitude and shape is predicted correctly by the simulation, with $\lambda_{Q^*}\approx 5~\mathrm{\mu m}$, and a broad maximum at $V_\mathrm{inj}\approx 400~\mathrm{\mu V}$. However, in detail the agreement is rather poor. The slope is too steep, both below and above the maximum. The relaxation rate due to impurity scattering has been calculated both within the quasiparticle approach,\cite{tinkham1972b,chi1979} and using quasiclassical Green's functions.\cite{takane2006} In contrast to the quasiparticle approach, the Green's function approach predicts different energy dependences for the clean and dirty limits. This difference might explain the discrepancy at low bias. At high bias, electron-phonon scattering in the Debye approximation\cite{kaplan1976} apparently overestimates the energy dependence of the relaxation rate. A weaker energy dependence has been predicted for electron-phonon scattering in the presence of disorder.\cite{bergmann1971,schmid1973,belitz1987} Also, it has been argued that disorder-enhanced electron-electron interaction may dominate in particular in aluminum with its weak electron-phonon interaction.\cite{stuivinga1983,clarke1986,lee1989} A simulation based on microscopic theory for both electron-electron and electron-phonon interaction might provide detailed insight into the relaxation mechanisms here, but this is beyond the scope of this article.

We now focus on the  dependence on magnetic field. Magnetic pair breaking adds an elastic contribution to the relaxation rate.\cite{schmid1975,artemenko1978,lemberger1981,lemberger1981b,takane2008} Consequently, as $B$ increases, the initial slope of the differential conductance decreases, and the maximum, i.e., the transition to the two-stage relaxation regime, shifts to higher bias. We find that the magnetic contribution to the relaxation rate is directly proportional to the pair-breaking parameter $\Gamma$, and that its energy dependence follows the theoretical prediction independent of sample details. Our observation $\tau_{Q^*}^{-1}\propto \Gamma$ is markedly different from the well-established approximation $\tau_{Q^*}^{-1}\propto \sqrt{1+2\tau_\mathrm{E}\Gamma/\hbar}$ valid for $T\rightarrow T_\mathrm{c}$,\cite{schmid1975} where $\tau_\mathrm{E}$ is the energy relaxation time (note that our $\Gamma/\hbar$ is the magnetic pair-breaking rate, i.e., the quantity $\tau_\mathrm{s}^{-1}$ of Ref.~\onlinecite{schmid1975}). Magnetic relaxation dominates all other contributions even at very low magnetic fields, where the spectral properties of the superconductor are almost unaffected by pair breaking. 

We finally discuss the possible impact of charge imbalance on the observation of other phenomena. We first note that the non-local conductance at subgap energies remains negligible at lowest temperatures, even with an applied magnetic field. This is not surprising, since even in the presence of magnetic pair breaking the superconductor still has a well-defined energy gap for quasiparticle excitations, at least as long as pair breaking is not too strong. Also, the relaxation length is always larger than the coherence length ($\xi\approx 100~\mathrm{nm}$ for our samples). Thus, the dependence of non-local conductance on bias or contact distance remains a good criterion to distinguish coherent subgap transport from charge imbalance. On the other hand, the observation of spin-dependent quasiparticle transport necessarily involves injection at energies above the gap, and ferromagnetic electrodes must be used. At magnetic fields of $\sim 100~\mathrm{mT}$, which are easily reached by the fringing fields of electrodes made of elementary ferromagnets, $\lambda_{Q^*}$ can already be as small as $1~\mathrm{\mu m}$. This is similar to the spin-diffusion length $\lambda_\mathrm{sf}$ in aluminium,\cite{jedema2002,jedema2003} and consequently great care must be taken to distinguish charge imbalance and spin-dependent transport. 

\section{Conclusion}

In conclusion, we have presented a detailed investigation of charge imbalance in superconductors in the low-temperature regime. From our measurements, we have extracted the charge-imbalance relaxation length as a function of bias voltage, temperature, and magnetic field. The bias-dependent results allow for a detailed comparison with different relaxation mechanisms. In particular, we have shown a transition from dominant elastic relaxation in the vicinity of the energy gap to an inelastic two-stage relaxation at high bias. The dependence on magnetic field follows theory with remarkable accuracy, and is clearly different from the known approximations for the high-temperature regime. The strong reduction of the relaxation length with magnetic field has possible implications for the interpretation of spin-diffusion experiments. 

During the preparation of this manuscript we became aware of three related studies of charge imbalance at low temperatures.\cite{tsuboi2010,kleine2010,arutyunov2010} We thank K.~Yu.~Arutyunov, W.~Belzig and D.~S.~Golubev for useful discussions. This work was supported by the DFG Center for Functional Nanostructures, and by the DAAD within the RISE program.

\bibliography{../../../lit.bib}

\end{document}